\documentstyle[12pt]{article}
\begin{document}
\thispagestyle{empty}
\title{Nonlinear Spinor Field in Bianchi type-I Universe filled with 
Perfect Fluid: Exact Self-consistent Solutions}
\author{B. Saha\\ 
Laboratory of Theoretical Physics\\ 
Joint Institute for Nuclear Research, Dubna\\ 
141980 Dubna, Moscow region, Russia\\ 
e-mail:  saha@thsun1.jinr.dubna.su\\ 
G. N. Shikin\\ 
Department of Theoretical Physics\\ 
Russian Peoples' Friendship University\\ 
6, Miklukho-Maklay str., 117198 Moscow, Russia}
\date{}
\maketitle 
\thispagestyle{empty}

\noindent
Self-consistent solutions to nonlinear spinor field  equations 
in General Relativity have been studied for the case of Bianchi type-I 
space-time filled with perfect fluid. The initial and the asymptotic
behavior of the field functions and the metric one has been thoroughly  
studied. It  should  be  emphasized  the  absence  of  initial 
singularity for some types of solutions and  also  the  isotropic 
mode of space-time expansion in some special cases.


\newpage
\section{Introduction}

\noindent
The quantum field theory in curved space-time has been a matter of
great interest in recent years because of its applications to 
cosmology and astrophysics. The evidence of existence of strong
gravitational fields in our Universe led to the study of the quantum 
effects of material fields in external classical gravitational field.
After the appearance of Parker's paper on scalar fields \cite{Par1} and 
spin-$\frac{1}{2}$ fields \cite{Par2}, several authors have studied this 
subject. Although the Universe seems homogenous and isotropic at present,
there are no observational data guarantying the isotropy in the era prior
to the recombination. In fact, there are theoretical arguments that sustain 
the existence of an anisotropic phase that approaches an isotropic one 
\cite{Mis}. Interest in studying Klein-Gordon and Dirac equations in
anisotropic models has increased since Hu and Parker \cite{Hu0} have shown
that the creation of scalar particles in anisotropic backgrounds can
dissipate the anisotropy as the Universe expands. 

\noindent
A Bianchi type-I (B-I) Universe, being the straightforward generalization 
of the flat Robertson-Walker (RW) Universe, is one of the simplest models  
of an anisotropic Universe that describes a homogenous and spatially flat
Universe. Unlike the RW Universe which has the same 
scale factor for each of the three spatial directions, a B-I Universe
has a different scale factor in each direction, thereby introducing an
anisotropy to the system. It moreover has the agreeable property that
near the singularity it behaves like a Kasner Universe even in the 
presence of matter and consequently falls within the general analysis
of the singularity given by Belinskii et al \cite{Bel}. 
And in a Universe filled with matter for $p\,=\,\gamma\,\varepsilon, \quad 
\gamma < 1$, it has been shown that any initial anisotropy in a B-I
Universe quickly dies away and a B-I Universe eventually evolve
into a RW Universe \cite{Jac}. Since the present-day Universe is 
surprisingly isotropic, this feature of the B-I Universe makes it a prime 
candidate for studying the possible effects of an anisotropy in the early 
Universe on present-day observations. In light of the importance of
mentioned above, several authors have studied linear spinor field equations       
\cite{Chim}, \cite{Cas} and the behavior of 
gravitational waves (GW's) \cite{Hu}, \cite{Mied}, \cite{Cho}    
in B-I Universe. Nonlinear spinor field (NLSF) in external cosmological
gravitation field was first studied by G. N. Shikin in 1991  
\cite{Shik}. This study was extended by us for more general case where 
we consider nonlinear term as an arbitrary function of all possible 
invariants generated from spinor bilinear forms. In that paper we also
studied the possibility of elimination of initial singularity specially for
Kasner Universe \cite{Ryb1}. 
In a recent paper \cite{Ryb} we studied 
the behavior of self-consistent NLSF in B-I Universe that was 
followed by the papers \cite{Alv}, \cite{AlvIz} where we studied
the self-consistent system of interacting spinor and scalar fields.
The purpose of the paper is to extend our study for more general NLSF
in presence of perfect fluid. In the section 2 we derive fundamental
equations corresponding to the Lagrangian for the self-consistent system
of spinor and gravitational fields in presence of perfect fluid and seek 
their general solutions. In section 3 we give a detail analysis of
the solutions obtained for different kinds of nonlinearity. In section 4
we study the role of perfect fluid and in section 5 we sum up the
results obtained.
\vskip 5mm
\section{Fundamental equations and general solutions}
\setcounter{equation}{0}
\noindent
The Lagrangian for the self-consistent system of  spinor  and gravitation 
fields in presence of perfect fluid is 
\begin{equation} L=\frac{R}{2\kappa}+\frac{i}{2} 
\biggl[ \bar \psi \gamma^{\mu} \nabla_{\mu} \psi- \nabla_{\mu} \bar 
\psi \gamma^{\mu} \psi \biggr] - m\bar \psi \psi + L_N +L_m,
\end{equation} 
with $R$ being the scalar curvature and $\kappa$ being  the  
Einstein's gravitational constant. The nonlinear term
$L_N$ describes the self-interaction of spinor field and can be presented
as some arbitrary functions of invariants generated from the real bilinear 
forms of spinor field having the form: 
$$S\,=\, \bar \psi \psi, \quad                    
P\,=\,i \bar \psi \gamma^5 \psi, \quad
v^\mu\,=\,(\bar \psi \gamma^\mu \psi), \quad
A^\mu\,=\,(\bar \psi \gamma^5 \gamma^\mu \psi), \quad
T^{\mu\nu}\,=\,(\bar \psi \sigma^{\mu\nu} \psi),$$
where $\sigma^{\mu\nu}\,=\,(i/2)[\gamma^\mu\gamma^\nu\,-\,
\gamma^\nu\gamma^\mu]$. Invariants, corresponding to the bilnear forms, look 
$$ I = S^2, \quad J = P^2, \quad 
I_v = v_\mu\,v^\mu\,=\,(\bar \psi \gamma^\mu \psi)\,g_{\mu\nu}
(\bar \psi \gamma^\nu \psi),$$ 
$$I_A = A_\mu\,A^\mu\,=\,(\bar \psi \gamma^5 \gamma^\mu \psi)\,g_{\mu\nu}
(\bar \psi \gamma^5 \gamma^\nu \psi), \quad
I_T = T_{\mu\nu}\,T^{\mu\nu}\,=\,(\bar \psi \sigma^{\mu\nu} \psi)\,
g_{\mu\alpha}g_{\nu\beta}(\bar \psi \sigma^{\alpha\beta} \psi).$$ 
According to the Pauli-Fierz theorem \cite{Ber} among the five invariants
only $I$ and $J$ are independent as all other can be expressed by them:
$I_v = - I_A = I + J$ and $I_T = I - J.$ Therefore we choose the nonlinear
term $L_N = F(I, J)$, thus claiming that it describes the nonlinearity
in the most general of its form. $L_m$ is the Lagrangian of 
perfect fluid.\\[2mm] 
\noindent
We choose B-I space-time metric in the form 
\begin{equation}
ds^2\,=\,dt^2 - \gamma_{ij}(t)\,dx^i\,dx^j.
\end{equation}
As it admits no rotational matter, the spatial metric $\gamma_{ij}(t)$
can be put into diagonal form. Now we can rewrite the B-I space-time metric 
in the form \cite{Zel}:
\begin{equation} ds^2 = dt^2 - a^{2}(t)\,dx^{2} - b^{2}(t)\,dy^{2} -
c^{2}(t)\,dz^2,
\end{equation}
where the velocity of light is taken to be unity.
Einstein equations for $a(t), b(t)$ and $c(t)$ 
corresponding to the metric (2.3) and Lagrangian (2.1) read~\cite{Zel}:  
\begin{eqnarray}
\frac{\ddot a}{a} +\frac{\dot a}{a} \biggl(\frac{\dot b}{b}+\frac{\dot 
c}{c}\biggr)= -\kappa \biggl(T_{1}^{1}- \frac{1}{2}T\biggr), \\
\frac{\ddot b}{b} +\frac{\dot b}{b} \biggl(\frac{\dot a}{a}+\frac{\dot 
c}{c}\biggr)= -\kappa \biggl(T_{2}^{2}- \frac{1}{2}T\biggr),  \\
\frac{\ddot c}{c} +\frac{\dot c}{c} \biggl(\frac{\dot a}{a}+\frac{\dot 
b}{b}\biggr)= -\kappa \biggl(T_{3}^{3}- \frac{1}{2}T\biggr),   \\
\frac{\ddot a}{a} +\frac{\ddot b}{b} +\frac{\ddot 
c}{c}= -\kappa \biggl(T_{0}^{0}- \frac{1}{2}T\biggr),
\end{eqnarray}
where points denote differentiation with respect to t, and 
$T=T_{\mu}^{\mu}.$ 

\noindent
NLSF equations  and  components  of  
energy-momentum tensor for the spinor field and perfect fluid 
corresponding to (2.1) are
\begin{eqnarray}
i\gamma^\mu \nabla_\mu \psi -m\psi + F_{I} 2 S \psi + F_{J} 2P i 
\gamma^5 \psi\,&=&\,0, \nonumber \\
i \nabla_\mu \bar \psi \gamma^\mu + m \bar\psi - F_{I} 2 S \bar \psi 
- F_{J} 2P i \bar \psi \gamma^5 \,&=&\,0, 
\end{eqnarray}
where $F_{I}:= \partial F/\partial I$ and 
$F_{J}:= \partial F/\partial J.$
\begin{equation}
T_{\mu}^{\rho}=\frac{i}{4} g^{\rho\nu} \biggl(\bar \psi \gamma_\mu 
\nabla_\nu \psi + \bar \psi \gamma_\nu \nabla_\mu \psi - \nabla_\mu \bar 
\psi \gamma_\nu \psi - \nabla_\nu \bar \psi \gamma_\mu \psi \biggr) \,-
\delta_{\mu}^{\rho}L_{sp}+ T_{\mu\,(m)}^{\rho},
\end{equation}
while $L_{sp}$ on account of spinor field equations takes the form:
$$ L_{sp}\,=\,-\biggl[\frac{1}{2}\biggl(\bar \psi \frac{\partial 
L_N}{\partial \bar \psi}+ \frac{\partial L_N}{\partial \psi} \psi 
\biggr)-L_N\biggr]\,= -\bigl[2 I\,F_{I}\, + 2 J\,
F_{J} - L_N\bigr].$$
Here $T_{\mu\,(m)}^{\rho}$ is the 
energy-momentum tensor of perfect fluid. For a Universe filled with
perfect fluid, in the concomitant system of reference 
$(u^0=1, \, u^i=0, i=1,2,3)$ we have
\begin{equation}
T_{\mu (m)}^{\nu}\,=\, (p + \varepsilon) u_\mu u^\nu - 
\delta_{\mu}^{\nu} p \,=\,(\varepsilon,\,- p,\,- p,\,- p),
\end{equation} 
where energy $\varepsilon$ is related to the pressure $p$ by the
equation of state $p\,=\,\gamma\,\varepsilon$, the general solution
has been derived by Jacobs \cite{Jac}. $\gamma$ varies between the
interval $0\,\le\, \gamma\,\le\,1$, whereas $\gamma\,=\,0$ describes
the dust Universe, $\gamma\,=\,\frac{1}{3}$ presents radiation Universe,
$\frac{1}{3}\,<\,\gamma\,<\,1$ ascribes hard Universe and $\gamma\,=\,1$
corresponds to the stiff matter.   
In (2.8) and (2.9) $\nabla_\mu$ denotes the covariant 
derivative of spinor, having the form \cite{Zelnor}:  
\begin{equation} \nabla_\mu 
\psi=\frac{\partial \psi}{\partial x^\mu} -\Gamma_\mu \psi, \end{equation} 
where $\Gamma_\mu(x)$ are spinor affine connection matrices.  
$\gamma^\mu(x)$ matrices are defined for the metric (2.3) as follows. 
Using the equalities  \cite{Brill}, \cite{Wein}
$$ g_{\mu \nu} (x)= e_{\mu}^{a}(x) e_{\nu}^{b}(x) \eta_{ab},
\qquad \gamma_\mu(x)\,=\,e_{\mu}^{a}(x)\bar\gamma^a,$$ 
where $\eta_{ab}= \mbox{diag}(1,-1,-1,-1)$,
$\bar \gamma_\alpha$ are the Dirac matrices of Minkowski space and
$e_{\mu}^{a}(x)$ are the set of tetradic 4-vectors, we obtain 
the Dirac matrices $\gamma^\mu(x)$ of curved space-time
$$ \gamma^0=\bar \gamma^0,\quad \gamma^1 =\bar \gamma^1 /a(t),\quad 
\gamma^2= \bar \gamma^2 /b(t),\quad \gamma^3 = \bar \gamma^3 /c(t), $$
$$ \gamma_0=\bar \gamma_0,\quad \gamma_1 =\bar \gamma_1 a(t),\quad 
\gamma_2= \bar \gamma_2 b(t),\quad \gamma_3 = \bar \gamma_3 c(t). $$
$\Gamma_\mu(x)$ matrices are defined by the equality $$\Gamma_\mu (x)= 
\frac{1}{4}g_{\rho\sigma}(x)\biggl(\partial_\mu e_{\delta}^{b}e_{b}^{\rho} 
- \Gamma_{\mu\delta}^{\rho}\biggr)\gamma^\sigma\gamma^\delta, $$ 
which gives
\begin{equation} \Gamma_0=0, \quad \Gamma_1=\frac{1}{2}\dot a(t) 
\bar \gamma^1 \bar \gamma^0, \quad \Gamma_2=\frac{1}{2}\dot b(t) \bar 
\gamma^2 \bar \gamma^0, \quad \Gamma_3=\frac{1}{2}\dot c(t) \bar \gamma^3 
\bar \gamma^0.\end{equation}
Flat space-time matrices we choose in the form, given in 
\cite{Bog}:
\begin{eqnarray}
\bar \gamma^0&=&\left(\begin{array}{cccc}1&0&0&0\\0&1&0&0\\
0&0&-1&0\\0&0&0&-1\end{array}\right), \quad
\bar \gamma^1\,=\,\left(\begin{array}{cccc}0&0&0&1\\0&0&1&0\\
0&-1&0&0\\-1&0&0&0\end{array}\right), \nonumber\\
\bar \gamma^2&=&\left(\begin{array}{cccc}0&0&0&-i\\0&0&i&0\\
0&i&0&0\\-i&0&0&0\end{array}\right), \quad
\bar \gamma^3\,=\,\left(\begin{array}{cccc}0&0&1&0\\0&0&0&-1\\
-1&0&0&0\\0&1&0&0\end{array}\right).  \nonumber
\end{eqnarray}
Defining $\gamma^5$ as follows
\begin{eqnarray}
\gamma^5&=&-\frac{i}{4} E_{\mu\nu\sigma\rho}\gamma^\mu\gamma^\nu
\gamma^\sigma\gamma^\rho, \quad E_{\mu\nu\sigma\rho}= \sqrt{-g}
\varepsilon_{\mu\nu\sigma\rho}, \quad \varepsilon_{0123}=1,\nonumber \\
\gamma^5&=&-i\sqrt{-g} \gamma^0 \gamma^1 \gamma^2 \gamma^3 
\,=\,-i\bar \gamma^0\bar \gamma^1\bar \gamma^2\bar \gamma^3 =
\bar \gamma^5, \nonumber
\end{eqnarray}
we obtain
\begin{eqnarray}
\bar \gamma^5&=&\left(\begin{array}{cccc}0&0&-1&0\\0&0&0&-1\\
-1&0&0&0\\0&-1&0&0\end{array}\right).\nonumber
\end{eqnarray}
We study the space-independent solutions to NLSF equation (2.8).
In this case the first equation of the system (2.8) together  with (2.11) 
and (2.12) is
\begin{equation} i\bar \gamma^0 
\biggl(\frac{\partial}{\partial t} +\frac{\dot \tau}{2 \tau} \biggr) \psi 
-m \psi +{\cal D} \psi + i {\cal G} \gamma^5\psi=0, \quad  
\tau(t)=a(t)b(t)c(t),
\end{equation}
where we denote ${\cal D} := \, 2 S\, F_{I}$ and
${\cal G}:=\,  2 P\,F_{J}.$ For the components 
$\psi_\rho= V_\rho(t)$, where $\rho=1,2,3,4,$ from (2.13) one deduces
the following system of equations:  
\begin{eqnarray} 
{\dot V}_1 +\frac{\dot \tau}{2 \tau} V_1 
+i(m- {\cal D}) V_1 - {\cal G}V_3 &=& 0,  \nonumber\\
{\dot V}_2 +\frac{\dot \tau}{2 \tau} V_2 
+i(m- {\cal D}) V_2 - {\cal G}V_4 &=& 0,  \nonumber\\
{\dot V}_3 +\frac{\dot \tau}{2 \tau} V_3 
-i(m- {\cal D}) V_3 + {\cal G}V_1 &=& 0,  \nonumber \\
{\dot V}_4 +\frac{\dot \tau}{2 \tau} V_4 
-i(m- {\cal D}) V_4 + {\cal G}V_2 &=& 0.
\end{eqnarray}

\noindent
Let us now define the equations for 
\begin{eqnarray}
P\,=\,i(V_1 V_{3}^{*}-V_{1}^{*}V_3 +V_2V_{4}^{*}-V_{2}^{*}V_4), \nonumber \\
R\,=\,(V_1 V_{3}^{*}+V_{1}^{*}V_3 +V_2V_{4}^{*}+V_{2}^{*}V_4), \nonumber \\
S\,=\,(V_{1}^{*} V_{1}+V_{2}^{*}V_2 -V_{3}^{*}V_{3}-V_{4}^{*}V_4). 
\end{eqnarray}
After a little manipulation one finds
\begin{eqnarray}
\frac{d S_0}{d t} -2 {\cal G}\, R_0\,=\,0, \nonumber\\
\frac{d R_0}{d t}+2 (m- {\cal D})\, P_0 + 2 {\cal G} S_0\,=\,0, \nonumber\\ 
\frac{d P_0}{d t}-2 (m- {\cal D})\, R_0\,=\,0,
\end{eqnarray}
where $S_0 = \tau S, \quad P_0 = \tau P, \quad R_0 = \tau R$.
From this system we obtain
\begin{eqnarray}
S_0 {\dot S}_0 + R_0 {\dot R}_0 +P_0 {\dot P}_0\,=\,0, \nonumber
\end{eqnarray}
that gives
\begin{equation}
S^2 + R^2 + P^2 \,=\, C^2/ \tau^2, \qquad C^2 = \mbox{const.}
\end{equation}
 
\noindent
Let us go back to the system of equations (2.14). It can be written as
follows if one defines $W_\alpha\,=\,\sqrt{\tau}\,V_\alpha$:
\begin{eqnarray} 
{\dot W}_1 +i\Phi W_1 - {\cal G}W_3 &=& 0, \quad
{\dot W}_2 +i\Phi W_2 - {\cal G}W_4 = 0, \nonumber\\
{\dot W}_3 -i\Phi W_3 + {\cal G}W_1 &=& 0, \quad
{\dot W}_4 -i\Phi W_4 + {\cal G}W_2 = 0,
\end{eqnarray} 
where $\Phi\,=\,m- {\cal D}$. Defining  $U(\sigma) = W (t)$, where 
$\sigma = \int\,{\cal G} dt$, we rewrite the foregoing system as:
\begin{eqnarray} 
U_{1}^{\prime} + i (\Phi/{\cal G}) U_{1} - U_{3} &=& 0, \qquad
U_{2}^{\prime} + i (\Phi/{\cal G}) U_{2} - U_{4} = 0, \nonumber\\
U_{3}^{\prime} - i (\Phi/{\cal G}) U_{3} + U_{1} &=& 0, \qquad
U_{4}^{\prime} - i (\Phi/{\cal G}) U_{4} + U_{2} = 0,
\end{eqnarray} 
where  prime ($^\prime$) denotes differentiation with respect to $\sigma$.
 
\noindent
Let us now solve the Einstein equations. To do it we first write
the expressions for the components of the energy-momentum tensor 
explicitly. Using the property of flat space-time Dirac matrices
and the explicit form of covariant derivative $\nabla_\mu$ one can  
easily find
\begin{equation}
T_{0}^{0}= m\,S \,-\,F(I,\,J) + \varepsilon, \quad 
T_{1}^{1}=T_{2}^{2}=T_{3}^{3}= 2 I\,F_{I}\,+\,2 J\,F_{J} 
- F(I,\,J) -\,p. 
\end{equation}
Summation of Einstein equations (2.4), (2.5) and (2.6) leads to the equation 
\begin{equation}
\frac{\ddot 
\tau}{\tau}=-\kappa(T_{1}^{1}+T_{2}^{2}+T_{3}^{3}-\frac{3}{2}T)=
\frac{3\kappa}{2}\,\bigl(mS +2 I\,F_{I}\,+\,2 J\,F_{J} \,-\,2\,
F(I,\,J) \,+\varepsilon - \,p\bigr).
\end{equation} 
In case if the right hand side of (2.21) be the 
function of $\tau(t)\,=\,a(t)b(t)c(t)$, this equation takes the form 
\begin{equation}
\ddot \tau+\Phi(\tau)=0.            
\end{equation}
As is known this equation possesses exact solutions  for  
arbitrary  function $\Phi(\tau)$.  Giving the explicit 
form of $L_N \,=\,F(I,\,J)$, from (2.21)  one  can find concrete 
function $\tau(t)=abc$. Once the value of $\tau$ is obtained, one can 
get expressions for components $V_\alpha(t), \quad \alpha =1,2,3,4.$
Let us express $a, b, c$ through $\tau$. For this we notice that
subtraction of Einstein equations  (2.4) - (2.5)  leads  to  
the equation 
\begin{equation}
\frac{\ddot a}{a}-\frac{\ddot b}{b}+\frac{\dot a \dot c}{ac}- 
\frac{\dot b \dot c}{bc}= \frac{d}{dt}\biggl(\frac{\dot a}{a}- 
\frac{\dot b}{b}\biggr)+\biggl(\frac{\dot a}{a}- \frac{\dot b}{b} \biggr) 
\biggl (\frac{\dot a}{a}+\frac{\dot b}{b}+ \frac{\dot c}{c}\biggr)= 0. 
\end{equation} 
Equation (2.23) possesses the solution
\begin{equation}
\frac{a}{b}= D_1 \mbox{exp} \biggl(X_1 \int \frac{dt}{\tau}\biggr), \quad 
D_1=\mbox{const.}, \quad X_1= \mbox{const.} \end{equation}
Subtracting equations (2.4) - (2.6) and (2.5) - (2.6) one finds   
the equations similar to (2.23), having solutions \begin{equation} 
\frac{a}{c}= D_2 \mbox{exp} \biggl(X_2 \int \frac{dt}{\tau}\biggr), \quad 
\frac{b}{c}= D_3 \mbox{exp} \biggl(X_3 \int \frac{dt}{\tau}\biggr),  
\end{equation}
where $D_2, D_3, X_2, X_3 $ are integration constants. There is a 
functional dependence between the constants 
$D_1,\, D_2,\, D_3,\, X_1,\, X_2,\, X_3 $:  
$$ D_2=D_1\, D_3, \qquad X_2= X_1\,+\,X_3.$$
Using the equations (2.24) and (2.25), we rewrite $a(t), b(t), c(t)$ in 
the explicit form:  
\begin{eqnarray} a(t) &=& 
(D_{1}^{2}D_{3})^{\frac{1}{3}}\tau^{\frac{1}{3}}\mbox{exp}\biggl[\frac{2X_1 
+X_3}{3} \int\,\frac{dt}{\tau (t)} \biggr], \nonumber \\
b(t) &=& 
(D_{1}^{-1}D_{3})^{\frac{1}{3}}\tau^{\frac{1}{3}}\mbox{exp}\biggl[-\frac{X_1 
-X_3}{3} \int\,\frac{dt}{\tau (t)} \biggr], \nonumber \\
c(t) &=& 
(D_{1}D_{3}^{2})^{-\frac{1}{3}}\tau^{\frac{1}{3}}\mbox{exp}\biggl[-\frac{X_1 
+2X_3}{3} \int\,\frac{dt}{\tau (t)} \biggr]. 
\end{eqnarray}
Thus the previous system of Einstein equations is completely integrated.  
In this process of integration only first three of the complete system of 
Einstein equations have been used. General solutions to these three second 
order equations have been obtained. The  solutions  contain  six arbitrary 
constants: $D_1, D_3, X_1, X_3 $  and two others, 
that  were obtained while solving  equation  (2.22).  
Equation  (2.7)  is  the consequence of first three of Einstein equations.  
To  verify  the correctness of obtained solutions, it is necessary to put 
$a, b, c$ into (2.7). It should lead either to identity or to  some  
additional constraint between the constants. Putting $a, b, c$  from  (2.26) 
into (2.7) one can get the following equality:  
\begin{equation}
\frac{1}{3 \tau}\biggl[3 \ddot \tau - 2\frac{\dot \tau^2}{\tau}+ 
\frac{2}{3 \tau}\biggl(X_{1}^{2}+X_1 X_3 +X_{3}^{2}\biggr)\biggr] = 
-\kappa \biggl(T_{0}^{0}-\frac{1}{2}T\biggr), 
\end{equation}
that guaranties the correctness of the solutions obtained.
In fact we can rewrite  (2.21) and (2.27) as
\begin{equation}
\frac{\ddot \tau}{\tau} \, = \, \frac{3 \kappa}{2} \,(T_{0}^{0} + T_{1}^{1}),
\end{equation}
and \begin{equation}
\frac{\ddot \tau}{\tau} - \frac{2}{3}\,\frac{\dot{\tau}^2}{\tau^2} +
\frac{2}{9 \tau^2} {\cal X} \, = \, - \frac{\kappa}{2}\, (T_{0}^{0} -
3 T_{1}^{1}),
\end{equation}
where ${\cal X} := X_{1}^{2} + X_1 X_3 + X_{3}^{2}$. Combining (2.28) and 
(2.29) together one gets the solution for $\tau$ in quadrature:
\begin{equation}
\int\,\frac{d\tau}{\sqrt{3\kappa \tau^2 T_{0}^{0} + {\cal X}/3}}\,=\,t.
\end{equation}  
Let us note that in our further study we exploit the equations (2.21)
to obtain $\tau$ and (2.27) to estimate integration constants.

\noindent
It should be emphasized that we are dealing with cosmological problem
and our main goal is to investigate the initial and the asymptotic behavior
of the field functions and the metric ones. As one sees, all these functions
are in some functional dependence with $\tau$: $\psi \sim 1/\sqrt{\tau}$
and $a_i \sim \tau^{1/3} e^{\pm \int dt/\tau}$. Therefore in our further
investigation we mainly look for $\tau$, though in some particular cases
we write down field and metric functions explicitly.
\vskip 5mm
\section{Analysis of the solutions obtained for some special choice
of nonlinearity}
\setcounter{equation}{0}
\noindent
Let us now study the system for some special choice of $L_N$. First we 
analyze the system only for the NLSF which will be followed by the study 
when the Universe is filled with perfect fluid. But first of all  
we study the linear case. The reason to get the solution to 
the self-consistent system of equations for the linear spinor and 
gravitational fields is the necessity of comparing this solution with that 
for the  system  of equations   for the nonlinear spinor   and 
gravitational  fields  that permits  to clarify the  role of nonlinear 
spinor terms in the evolution of the cosmological model in question. 
Using the equation (2.21) one gets
\begin{equation}
\tau(t)\,=\,(1/2)\,Mt^2\,+\,y_1t\,+\,y_0     
\end{equation}       
where $M=\frac{3}{2}\kappa mC_0,$\,\,$C_0 = C_{1}^{2}+C_{2}^{2}-C_{3}^{2}
-C_{4}^{2}$ and $y_1, y_0$ are the constants.  
In this case we get explicit expressions for the components of spinor
field functions and metric functions:
\begin{equation}
V_r(t)=(C_r/\sqrt{\tau})\,e^{-imt},\quad r = 1,2; \quad
V_l(t)=(C_l/\sqrt{\tau})\,e^{imt}, \quad l = 3,4.
\end{equation}
\begin{eqnarray} a(t) &=& 
(D_{1}^{2}D_{3})^{\frac{1}{3}}(\frac{1}{2}Mt^2+y_1t+y_0)^{\frac{1}{3}}
Z^{2(2X_1+X_3)/3B}, \nonumber \\ 
b(t) &=& 
(D_{1}^{-1}D_{3})^{\frac{1}{3}}(\frac{1}{2}Mt^2+y_1t+y_0)^{\frac{1}{3}}
Z^{-2(X_1-X_3)/3B}, \nonumber \\ 
c(t) &=& 
(D_{1}D_{3}^{2})^{-\frac{1}{3}}(\frac{1}{2}Mt^2+y_1t+y_0)^{\frac{1}{3}}
Z^{-2(X_1+2X_3)/3B}, 
\end{eqnarray}
where $Z = \frac{(t-t_1)}{(t-t_2)},\,\, B = M(t_1 -t_2),$ 
and $t_{1,2} = -y_1/M \pm \sqrt{(y_1/M)^2 - 2y_0/M}$\, are the 
roots  of the quadratic equation \, $Mt^2+2y_1t+2y_0 = 0.$ 
Substituting $\tau(t)$ into (2.27), one gets  
\begin{equation}
y_{1}^{2}- 2My_0\,=\,(X_{1}^{2}+X_1X_3+X_{3}^{2})/3\,=\,{\cal X}/3 \,>\,0.
\end{equation} 
This means that the quadratic polynomial in (3.1) possesses real roots, 
i.e. $\tau(t)$ in (3.1) turns into zero at $t=t_{1,2}$ and the solution 
obtained is the singular one. Let us now study the solutions (3.1) - (3.3) 
at $t \to \infty$. In this case we have 
$$\tau(t) \approx \frac{3}{4}\kappa mC_0 t^2, \qquad
a(t) \approx b(t) \approx c(t) \approx t^{2/3}, $$
that leads to the conclusion about the asymptotical 
isotropization of the expansion process for the initially anisotropic 
B-I space.  Thus the solution to the self-consistent system of  
equations for the linear spinor and gravitational fields  is  the singular 
one at the initial time. In the initial state of evolution of the field 
system the expansion process of space is  anisotropic,  but at $t \to 
\infty$ there happens isotropization of the expansion process. 

\noindent
Once the solutions to the linear spinor field equations and corresponding
to them metric functions are obtained, let us now study the nonlinear case.  
\vskip 3mm
\noindent
{\bf I.} Let us consider the case when $L_N\,=\,F(I)$.
It is clear that in this case ${\cal G}\,=\,0$.
From (2.16) we find  
\begin{equation}
S = C_0/\tau, \quad C_0= \mbox{const.}
\end{equation}
As in the considered case $L_N\,=\,F$ depends only on $S$, from (3.5) it
follows that $F(I)$ and $F_I(I)$ are functions of $\tau= 
abc$.  Taking this fact into account, integration  of the system of 
equations (2.14) leads to the expressions 
\begin{eqnarray} 
V_{r}(t) = (C_r/\sqrt{\tau})\,e^{-i\Omega}, \quad r=1,2, \quad 
V_{l}(t) = (C_l/\sqrt{\tau})\,e^{i\Omega}, \quad l=3,4.  
\end{eqnarray} where $ C_r$ and 
$C_l$ are integration constants. Putting (3.6) into (2.15) one gets
\begin{equation}
S = (C_{1}^{2}+C_{2}^{2}-C_{3}^{2}-C_{4}^{2})/\tau.
\end{equation}
Comparison of (3.5) with (3.7) gives $C_0 = 
C_{1}^{2}+C_{2}^{2}-C_{3}^{2}-C_{4}^{2}.$

\noindent
Let us consider the concrete type of NLSF  
equation with $F(I) =\lambda I^{(n/2)}=\lambda S^n$ where $\lambda$  
is the coupling constant, $n>1$. 
In this case for $\tau$ one gets: 
\begin{equation}
\ddot \tau = (3/2)\kappa C_0 \bigl[m+ \lambda (n-2) 
C_{0}^{n-1}/\tau^{n-1}\bigr].  
\end{equation}
The first integral of the foregoing equation takes form:
\begin{equation}
\dot \tau^2 = 3\kappa C_0 \bigl[m\tau - \lambda 
\, C_{0}^{n-1}/\tau^{n-2} + g^2\bigr], 
\end{equation} 
where from (2.27) one determines $g^2\,=\, {\cal X}/9\kappa\,C_0$.
The sign $C_0$  is determined by the positivity 
of the energy-density $T_{0}^{0}$  of linear spinor field:
\begin{equation}
T_{0}^{0} = m C_0/\tau > 0.
\end{equation}   
It is obvious from (3.10) that $C_0 >0.$ Now one can write the
solution to the equation  (3.9) in quadratures:
\begin{equation}
\int \frac{\tau^{(n-2)/2}d\tau}{\sqrt{m \tau^{n-1} +g^2 
\tau^{n-2}-\lambda C_{0}^{n-1}}}= \sqrt{3\kappa C_0}\,t 
\end{equation} 
The constant of integration in (3.11) has been taken zero, as it only gives 
the shift of the initial time.  Let us study the properties of solution to 
equation (3.8) for $n>2$. From (3.11) one gets 
\begin{equation}
 \tau(t)\mid_{t \to \infty} \approx (3/4) \kappa mC_0t^2,
\end{equation}
which coincides with the asymptotic solution to the equation (3.3). It leads  
to the conclusion about isotropization of the expansion process of the 
B-I space. It should be remarked that the isotropization takes place if and
only if the spinor field equation contains the massive term  [cf.  
the parameter m in (3.12)].  If m=0 the isotropization does not take place. 
In this case from (3.11) we get 
\begin{equation}
\tau(t)\mid_{t \to \infty} \approx \sqrt{3\kappa C_0 g^2}\,t.
\end{equation} 
Substituting (3.13) into (2.26) one comes  to  the conclusion  that the 
functions $a(t), b(t)$ and $c(t)$ are different.  Let us consider the 
properties of solutions to equation  (3.8) when $t \to 0.$ For $\lambda<0$ 
from (3.11) we get 
\begin{equation}
\tau(t)= \bigl[(3/4) n^2 \kappa |\lambda| 
C_{0}^{n}\bigr]^{1/n}t^{2/n} \to 0, \end{equation} 
i.e. solutions are singular. For $\lambda>0,$ from (3.11) it follows that 
$\tau=0$ cannot be reached for any value of $t$ as in this case the 
denominator of the integrand in (3.11) becomes imaginary. It means that 
for $\lambda>0$ there exist regular solutions to the previous system of 
equations \cite{Ryb}. The absence of the initial singularity in the 
considered cosmological solution appears to be consistent with the 
violation for $\lambda>0$, of the dominant energy condition in the 
Hawking-Penrose theorem \cite{Zel}.  

\noindent
Let us consider the Heisenberg-Ivanenko equation when in (3.8) n=2 
\cite{Ivan}. In this case  the equation  for $\tau(t)$ does not contain 
the nonlinear term and its solution coincides with that of the linear 
equation (3.3). With such $n$ chosen the metric functions $a, b, c$ are given by 
the equality (3.2), and the spinor field functions are written as follows:  
\begin{equation}
V_r = (C_r/\sqrt{\tau})\,e^{-imt}Z^{4i\lambda C_0/B}, \quad
V_l = (C_l/\sqrt{\tau})\,e^{imt}Z^{-4i\lambda C_0/B} 
\end{equation}
As in the linear case, the obtained solution is singular at initial time 
and asymptotically isotropic as $t \to \infty$.

\noindent  
We now study the properties of solutions to equation (3.8) for 
$1<n<2.$ In this case it is convenient to present the solution (3.11) in the 
form:
\begin{equation}
\int \frac{d \tau}{\sqrt{m\tau -\lambda \tau^{2-n} 
C_{0}^{n-1}+g^2}}=\sqrt{3\kappa C_0}\,t 
\end{equation} 
As $t \to \infty$, from (3.16) we get the equality (3.12), 
leading to the isotropization of the expansion process.  If  $m=0$ and 
$\lambda>0,$ \quad $\tau(t)$ lies on the interval 
$$0 \le \tau(t) \le \bigl(g^2/\lambda C_{0}^{n-1}\bigr)^{1/(2-n)}.$$
If m=0 and $\lambda<0,$ the relation (3.16) at $t \to \infty$ leads to the 
equality:  
\begin{equation}
\tau(t) \approx \bigl[(3/4)n^2 \kappa |\lambda| C_{0}^{n} 
\bigr]^{1/n}t^{2/n}.
\end{equation}  
Substituting (3.17) into (2.26) and taking into account that at $t \to 
   \infty$
$$ \int \frac{dt}{\tau} \approx \frac{n(3\kappa |\lambda| 
n^2C_{0}^{n})^{1/n}}{(n-2)2^{2/n}}t^{-2/n+1} \to 0 $$
due to $-2/n+1 < 0,$ we obtain 
\begin{equation}
a(t) \sim b(t) \sim c(t) \sim 
[\tau(t)]^{1/3} \sim t^{2/3n} \to \infty.  
\end{equation} 
It means that the solution obtained tends to the isotropic one.  In 
this case the isotropization is provided not by the massive parameter, 
but by the degree $n$ in the term $L_N = \lambda S^n.$  
(3.16) implies 
\begin{equation}
\tau(t)\mid_{t \to 0} \approx \sqrt{3\kappa C_0 g^2}\,t \to 0,
\end{equation} 
which means the solution obtained is initially singular. Thus, for $1<n<2$ 
there exist only singular solutions at initial time. At $t \to \infty$ 
the isotropization of  the expansion process of B-I space 
takes place both for $m\not= 0$ and for $m=0.$  

\noindent
Let us finally study the 
properties of   the solution to the equation (3.8) for $0<n<1.$ In this 
case  we use the solution in the form (3.16). As now $2-n>1,$ then 
with the increasing of $\tau(t)$ in the denominator of the integrand in 
(3.16)  the second term $\lambda \tau^{2-n} C_{0}^{n-1}$ increases faster 
than the first one.  Therefore the solution describing the space expansion 
can be possible only for $\lambda<0.$  In this case at $t\to \infty$, 
for $m=0$ as well as for $m\not= 0,$ one can get the asymptotic 
representation (3.17) of the solution.  This solution,  as for the choice 
$1<n<2,$ provides asymptotically isotropic expansion of the B-I 
space.  For $t \to 0$ in this case we shall get only singular 
solution of the form (3.19). 
\vskip 3mm
\noindent
{\bf II.} We study the system when $L_N\,=\,F(J)$,
which means in the case considered ${\cal D}\,=\,0$. Let us 
note that, in the unified nonlinear spinor theory of Heisenberg the 
massive term remains absent, as according to Heisenberg, the particle 
mass should be obtained as a result of quantization of spinor prematter 
\cite{Hei}. In nonlinear generalization of classical field equations, 
the massive term does not possess the significance that it possesses 
in linear one, as by no means it defines total energy (or mass) of 
nonlinear field system. Thus without losing the generality we can 
consider massless spinor field putting $m\,=\,0$ that leads to 
$\Phi\,=\,0.$ This assumption metamorphoses (2.16) to get
\begin{equation}
P(t)\,=\,D_0/\tau, \,\, D_0=\,\mbox{const.}
\end{equation}
The system of equations (2.19) in this case reads
\begin{eqnarray} 
U_{1}^{\prime} -  U_{3} &=& 0, \qquad
U_{2}^{\prime} -  U_{4} = 0, \nonumber\\
U_{3}^{\prime} +  U_{1} &=& 0, \qquad
U_{4}^{\prime} +  U_{2} = 0.
\end{eqnarray} 
Differentiating the first equation of system (3.21) and taking into 
account the third one we get 
\begin{equation}
U_{1}^{\prime \prime} +U_{1} =\,0,
\end{equation}
which leads to the solution
\begin{equation}
U_1 = D_1 e^{i \sigma} + iD_3 e^{-i \sigma},\quad
U_3 =  i D_1 e^{i \sigma} + D_3 e^{-i \sigma}.
\end{equation}
Analogically for $U_2$ and $U_4$ one gets
\begin{equation}
U_2 = D_2 e^{i \sigma} + iD_4 e^{-i \sigma},\quad
U_4 = i D_2 e^{i \sigma} + D_4 e^{-i \sigma}, 
\end{equation}
where $D_i$ are the constants of integration.
Finally, we can write
\begin{eqnarray}
V_1&=&(1/\sqrt{\tau}) (D_1 e^{i \sigma} + iD_3 
e^{-i\sigma}), \quad
V_2 = (1/\sqrt{\tau}) (D_2 e^{i \sigma} + iD_4
e^{-i\sigma}),
\nonumber \\
V_3&=&(1/\sqrt{\tau}) (iD_1 e^{i \sigma} + D_3
e^{-i \sigma}), \quad
V_4 = (1/\sqrt{\tau}) (iD_2 e^{i \sigma} + D_4
e^{-i\sigma}).
\end{eqnarray} 
Putting (3.25) into the expressions (2.15) one finds
\begin{equation}
P=2\,(D_{1}^{2} + D_{2}^{2} - D_{3}^{2} -D_{4}^{2})/\tau.
\end{equation}
Comparison of (3.20) with (3.26) gives
$D_0=2\,(D_{1}^{2} + D_{2}^{2} - D_{3}^{2} -D_{4}^{2}).$

Let us now estimate $\tau$ using the equation
\begin{equation}
\ddot{\tau}/\tau\,=\,3 \kappa \,\lambda (n - 1) P^{2n},
\end{equation}
where we chose $L_N\,=\,\lambda P^{2n}$. Putting the value of $P$ into
(3.20) and integrating one gets
\begin{equation}
\dot{\tau}^2 \, = \,- 3\kappa\,\lambda D_{0}^{2n} \tau^{2 - 2n} + y^2,
\end{equation}
where $y^2$ is the integration constant and can be defined from (2.27):
$y^2\,=\,{\cal X}/3 > 0$. The solution to the equation (3.28) in quadrature 
reads
\begin{equation}
\int\,\frac{d\tau}{\sqrt{- 3 \kappa\lambda D_{0}^{2n}\tau^{2 - 2n} + y^2}} 
\, = \, t.
\end{equation}
Let us now analyze the solution obtained here. As one can see
the case $n = 1$ is the linear one.  In case of $\lambda < 0$ for
$n > 1$ i.e. $2 - 2n < 0$, we get
$$ \tau(t)\mid_{t \to 0} \approx [(\sqrt{3 \kappa|\lambda|}  
D_{0}^{n}n) t]^{1/n},$$
and
$$ \tau\mid_{t \to \infty} \approx \sqrt{3\kappa y^2} \,t, $$
it means that for the term $L_N$ considered with $\lambda < 0$ and
$n > 1$ the solution is initially singular and the space-time is
anisotropic at $t \to \infty.$ 
Let us now study it for $n < 1$. In 
this case we obtain
$$ \tau\mid_{t \to 0} \approx \sqrt{3\kappa y^2}\, t, $$
and
$$ \tau\mid_{t \to \infty} \approx 
[(\sqrt{3\kappa|\lambda|} D_{0}^{n} n) t]^{1/n}.$$
The solution is initially singular as in previous case, but as far as
$ 1/n > 1$, it provides asymptotically isotropic expansion of B-I space-time.
\vskip 3mm
\noindent
{\bf III.} In this case we study $L_N\,=\,F(I,\,J)$. Choosing 
\begin{equation}
L_N\,=\,F(K_{\pm}), \quad K_{+} = I + J = I_v = -I_A, \quad
K_{-} = I - J = I_T,
\end{equation}
in case of massless NLSF we find
$$
{\cal D}\,=\,2 S F_{K_{\pm}}, \quad
{\cal G}\,=\, \pm 2 P F_{K_{\pm}}, \quad F_{K_{\pm}} = dF/dK_{\pm}.
$$
Putting them into (2.16) we find
\begin{equation}
S_{0}^{2} \pm P_{0}^{2} = D_{\pm}.
\end{equation}
Choosing $F = \lambda K_{\pm}^{n}$ from (2.21) we get
\begin{equation}
\ddot{\tau}\,=\,3 \kappa \lambda (n - 1)\,D_{\pm}^{n}\,\tau^{1-2n},
\end{equation}
with the solution
\begin{equation}
\int\,\frac{\tau^{n-1} d \tau}{\sqrt{g^2\tau^{2n - 2} - 3\kappa\lambda 
D_{\pm}^{n}}}
\,=\, t,
\end{equation}
where $g^2\,=\,{\cal X}/3.$ Let us study the case with $\lambda < 0$.
For $n < 1$ from (3.33) one gets
\begin{equation}
\tau (t)\mid_{t \to 0} \approx g t \to 0,
\end{equation}
i.e. the solutions are initially singular, and  
\begin{equation}
\tau (t)\mid_{t \to \infty} \approx 
[\sqrt{(3\kappa |\lambda| D_{\pm}^{n})}t]^{1/n},
\end{equation}
which means that the anisotropy disappears as the Universe expands.
In case of $n > 1$ we get 
$$ \tau (t)\mid_{t \to 0} \approx t^{1/n} \to 0, $$
and
$$ \tau (t)\mid_{t \to \infty} \approx gt, $$
i.e. the solutions are initially singular and the metric functions 
$a(t), b(t), c(t)$ are different at $ t \to \infty$, i.e.
the isotropization process remains absent. For $\lambda > 0$ we 
get the solutions those are initially regular, but it violates the
dominant energy condition in Hawking-Penrose theorem \cite{Zel}.
Note that one comes to the analogical conclusion choosing 
$L_N\,=\,\lambda S^{2n}P^{2n}.$
\vskip 5mm
\section{Analysis of the results obtained when the B-I 
Universe is filled with perfect fluid}
\setcounter{equation}{0}

\noindent
Let us now analyze the system filled with perfect fluid.
Let us recall that the energy-momentum tensor of perfect fluid is 
\begin{equation}
T_{\mu (m)}^{\nu}\,=\, (p + \varepsilon) u_\mu u^\nu - 
\delta_{\mu}^{\nu} p\,=\,(\varepsilon, - p, - p, - p). 
\end{equation}
As we saw earlier the introduction of perfect fluid does not change
the field equations, thus leaving the solutions to the NLSF equations
externally unchanged. Changes in the solutions performed by perfect fluid
carried out through Einstein equations, namely through $\tau$. So, let us 
first see how the quantities $\varepsilon$ and $p$ connected with $\tau$.
In doing this we use the well-known equality $T_{\mu;\nu}^{\nu}\,=\,0$,
that leads to
\begin{equation}
\frac{d}{dt}(\tau \varepsilon) + {\dot \tau} p\,=\,0,
\end{equation}
with the solution
\begin{equation}
\mbox{ln} \tau\,=\,-\int\,\frac{d \varepsilon}{(\varepsilon + p)}.
\end{equation}
Recalling the equation of state 
$p\,=\,\xi \varepsilon,\,\, 0 \le \xi \le 1$ finally we get
\begin{equation}
T_{0 (m)}^{0}\,=\,\varepsilon\,=\,\frac{\varepsilon_0}{\tau^{1+\xi}}, \,\,
T_{1 (m)}^{1}\,=\,T_{2 (m)}^{2}\,=\,T_{3 (m)}^{3}\,=\,- p\,=\,-
\frac{\varepsilon_0 \xi}{\tau^{1+\xi}},
\end{equation}
where $\varepsilon_0$ is the integration constant. Putting them into (2.21)
we get 
\begin{equation}
\frac{\ddot \tau}{\tau}\,=\,\frac{3 \kappa}{2}\frac{(\xi - 1)
\varepsilon_0}{\tau^{(\xi + 1)}}
\end{equation}
which shows that for stiff matter $(\xi = 1)$ the contribution 
of fluid to the solution is missing. 
Let us now study the system with nonlinearity type {\bf I}. In this case
we get
\begin{equation}
\int\,\frac{d \tau}{\sqrt{m C_0 \tau - \lambda C_{0}^{n}/\tau^{(n-2)} +
\varepsilon_0 \tau^{(1 - \xi)} + g^2}}\,=\,\pm \sqrt{3 \kappa} t.
\end{equation}
As one can see in case of dust $(\xi = 0)$  the fluid term can be combined 
with the massive one, whereas in case of stiff matter $(\xi = 1)$
it mixes up with the constant. Analyzing the equation (4.6) one comes to the 
conclusion that the presence of perfect fluid does not influence the 
result obtained earlier for the nonlinear term type {\bf I}. One comes 
to the same conclusion analyzing the system with perfect fluid for the 
other types of nonlinear terms considered here. At least
both at $t \to 0$ and at $t \to \infty$ the key role is 
played by the other terms rather than the term
presenting fluid. 
\vskip 5mm
\section{Conclusions}

\noindent
Exact solutions to the NLSF equations have been obtained for the 
nonlinear terms being arbitrary functions of the invariant 
$I = S^2$ and $J = P^2$, where $S=\bar \psi \psi$ and 
$P= i \bar \psi \gamma^5 \psi$ are the real bilinear forms of spinor 
field, for B-I space-time. Equations with power nonlinearity in  
spinor field Lagrangian $L_N = \lambda S^n$, where $\lambda$ is the 
coupling constant, have been thoroughly studied. In this case it is shown 
that equations mentioned possess solutions both regular and singular at 
the initial moment of time for $n>2$ . Singularity remains  absent  for  
the case of field system with broken dominant energy condition. It is also 
shown that if in the NLSF equation the massive parameter 
$m \ne 0$ and $n\ge 2$ then at $t \to \infty$ isotropization of  
B-I space-time expansion takes place, while for $m=0$ the 
expansion is anisotropic. Properties of solutions to the spinor field 
equation for $1<n<2$ and $0<n<1$ we also studied. It was found that in 
these cases there does not exist solution that is initially regular. 
At $t \to \infty$ the isotropization process of B-I
space-time takes place both for $m \ne 0$ and for $m = 0$. In case of
nonlinear term $L_N = \lambda P^{2n}$, we found the solutions those are
initially singular and the isotropization process of B-I space-time
depends on the choice of $n$. For $L_N = \lambda (I \pm J)^{n}$ 
we obtained the solutions those may be initially singular or
regular depends on the sign of coupling constant $\lambda$, but 
isotropization process depends on the value of power $n$. It is also 
shown that the results remain unchanged even in the case when the B-I 
space-time is filled with perfect fluid.

\end{document}